\documentclass{ws-p10x7}

\begin{document}

\title{$D^0$-$\bar D^0$ Mixing and CP Violation from FOCUS Experiment}

\author{Hwanbae Park}

\address{Department of Physics, Korea University, Seoul, Korea \\
           {\sc   on behalf of the FOCUS(E831) Collaboration}
         }

\twocolumn[\maketitle\abstract{
Measurement results on $D^0$ - $\bar D^0$ mixing and CP violation are presented.
FOCUS, a fixed target experiment at Fermilab, collected  a high
statistics photo-produced charm sample during the 96-97 run.
We reconstructed more 
than 1 million charmed particles and compared the lifetimes of two
$D^0$ meson decays to
$K^-\pi^+$ and $K^- K^+$. 
The mixing parameter, $y_{\rm cp}$, we obtained is $(3.42\pm 1.39\pm 0.74)\%$.
We also searched CP asymmetries in
$D^+\rightarrow K^-K^+\pi^+$, $D^0\rightarrow K^-K^+$ and $D^0 \rightarrow\pi^-\pi^+$
decay modes. We did not see any evidence of CP violation by comparing the
decay rates for particle and antiparticle.
        }]

\section{Introduction}
 FOCUS\footnote{Coauthors: J.Link, V.S. Paolone, M. Reyes, P.M. Yager
({\bf UC DAVIS}); J.C. Anjos,
 I. Bediaga, C. G\"obel, J. Magnin, J.M. de Miranda, I.M. Pepe
% ...
}  ($\bf{\rm Pho}$toproduction of $\bf{\rm C}$harm with an $\bf{\rm U}$pgraded
 $\bf{\rm S}$pectrometer) is the successor of the E687 experiment~\cite{E687} with a significantly
upgraded spectrometer, and it is designed to study charm physics.
Charm particles are produced by the interaction of roughly 180 GeV high energy photons 
with a segmented beryllium oxide target.

During the 96$-$97 fixed target run at Fermilab, we
collected more than 6.3 billion events and reconstructed more than
1 million charm particles in $D\rightarrow K\pi$,$K2\pi$ and $K3\pi$ decay modes. 
From this sample, we measured the lifetime differences in the $D^0$ meson system and
searched for CP violation in singly Cabibbo suppressed D meson decays.
Since the Standard Model predictions for these processes are extremely small, any
observation of a signal could be a clear indication of physics beyond the Standard Model.

\section{$D^0 - \bar D^0$ mixing}
In hadronic decays of the neutral charm meson there is an interference term between the mixing
and the doubly Cabibbo suppressed paths. By direct comparison of a lifetime difference between
weak eigenstates we can search for charm mixing.
Assuming both CP is conserved, and $D^0\rightarrow K^-\pi^+$ is an equal mixture state of CP even and odd, then

\begin{equation}
y_{\rm cp}=\frac{\tau(D\rightarrow K\pi)}{\tau(D\rightarrow KK)}\! -1
\label{eq:app}
\end{equation}
where $D^0\rightarrow K^-K^+$ is a CP even eigenstate. 

For analysis the sample was selected by requiring either $D^*$ tagging ($\it tagged~sample$), which satisfy
the mass difference between $D^*$ and $D$ is within 3 $\rm {MeV/c^2}$ of the nominal value,
or requiring more stringent cuts on particle identification for kaons and pions, momentum
asymmetry ($\frac{|P_1 - P_2|}{P_1 + P_2}$) between two daughter tracks, 
the resolution in decay proper time and
requirement of primary vertex inside the target material ($\it inclusive~sample$).
The cuts are chosen not to bias the reduced proper time distribution.
The tagged sample has clean signals while the inclusive sample gives a larger
sample of events. After all of cuts applied, we have 119738 signal events in $K\pi$ and
10331 in $KK$ from the combination of two samples.
\begin{figure*}
\epsfxsize180pt
\figurebox{32pt}{80pt}{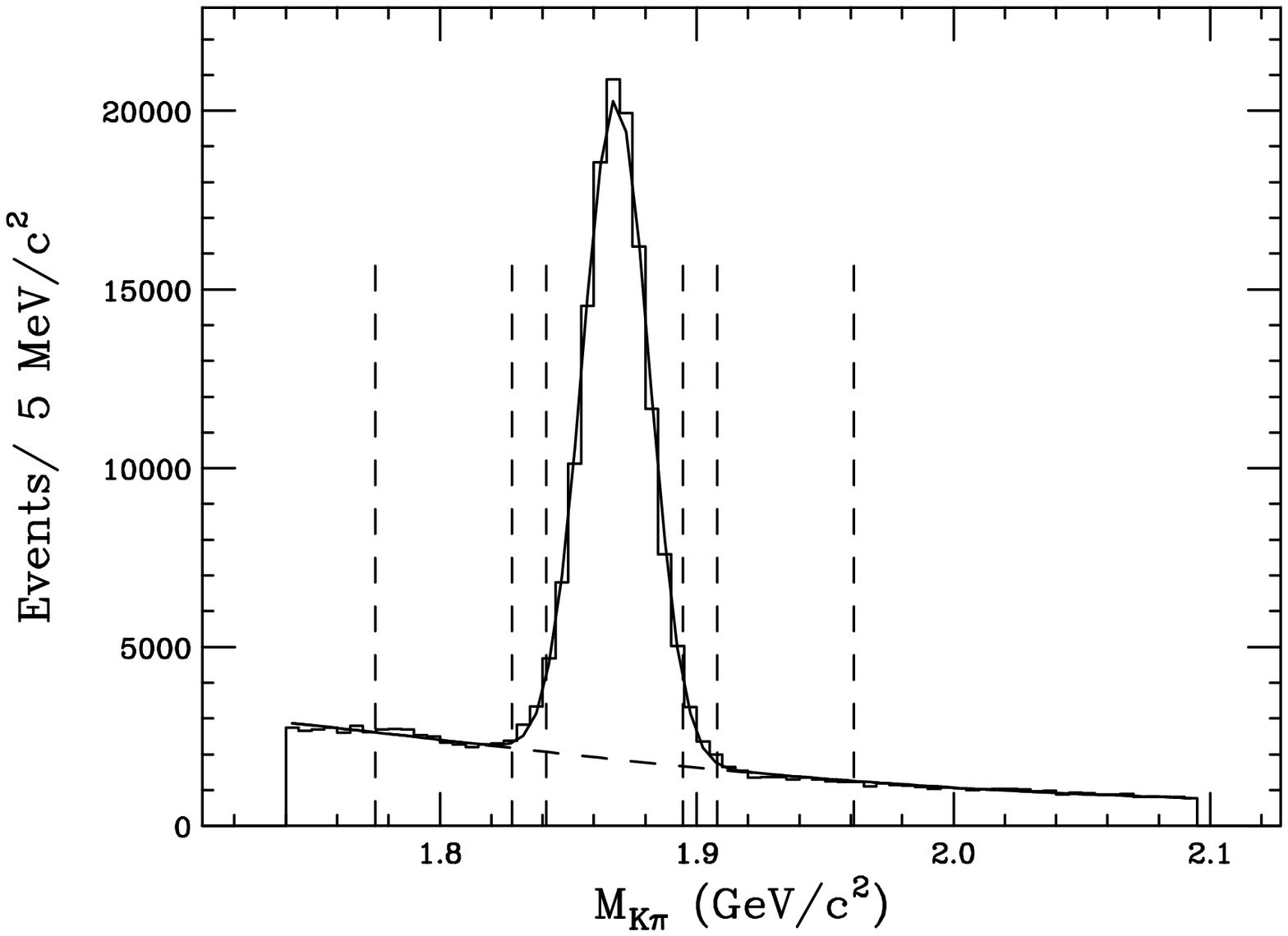}
\epsfxsize180pt
\figurebox{32pt}{80pt}{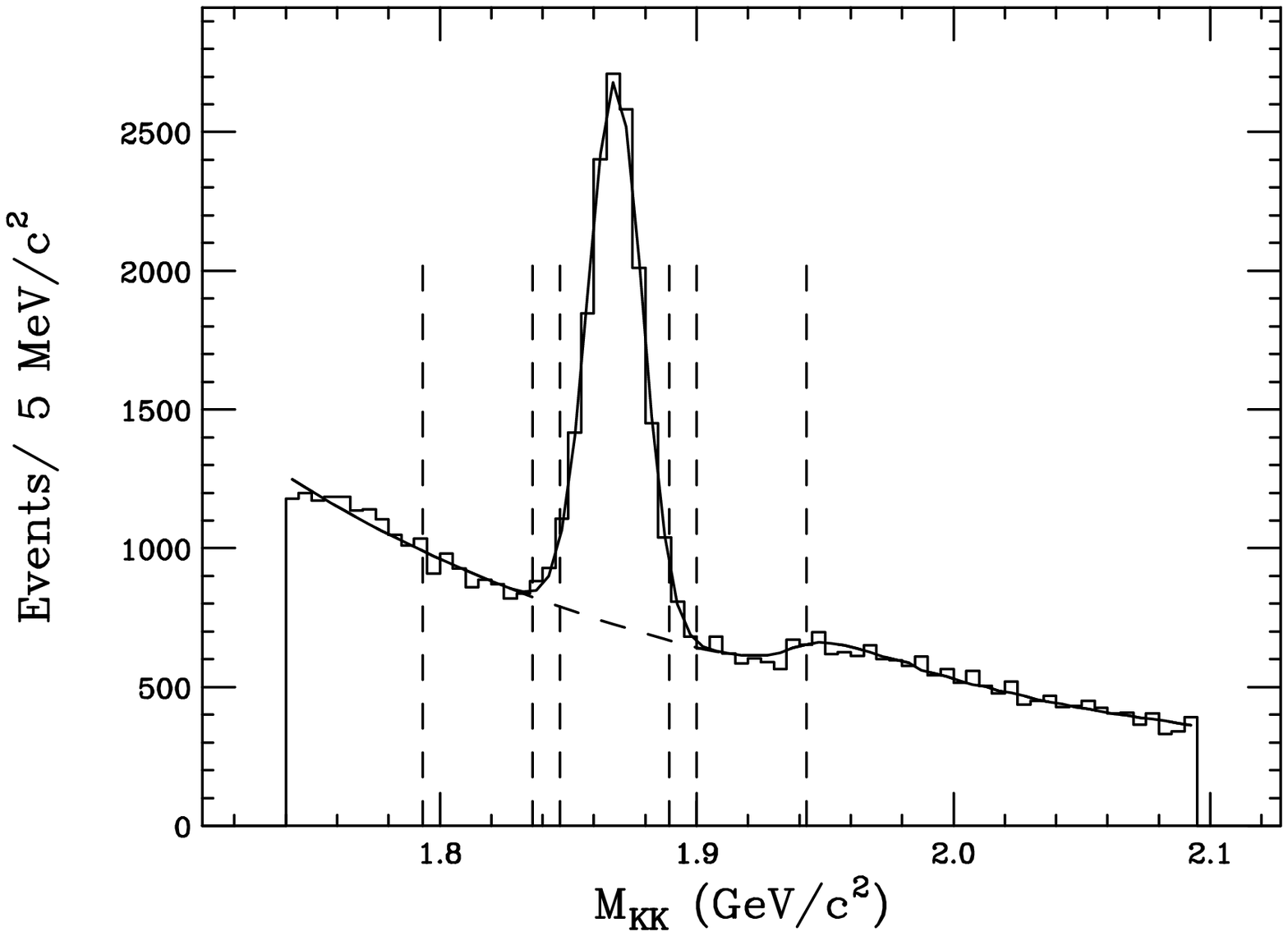}
\caption{Signal for $D^0 \rightarrow K^- \pi^+$ and $K^-K^+$ with 
a detachment cut of $\ell/\sigma > 5$. 
The reflection in the
background at higher masses in $KK$ is due to contamination from misidentified
$D^0 \rightarrow K^- \pi^+$.
The yields are 119738 and 10331 for $K^-\pi^+$ and $K^-K^+$ signal
events, respectively. 
The
vertical dashed lines indicate the signal and sideband regions used
for the lifetime and $y_{\rm cp}$ fits.}
\label{kkmass_lifetime}
\end{figure*}
\begin{figure}
\epsfxsize160pt
\figurebox{120pt}{160pt}{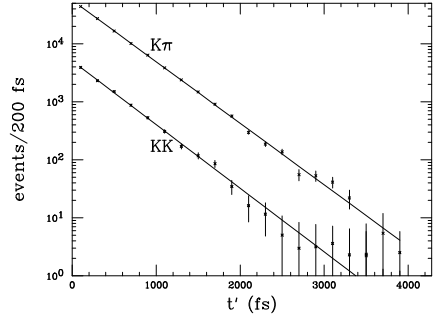}
\caption{ Signal versus reduced proper time for $D^0 \rightarrow K^-\pi^+~ ~{\rm and}~~K^- K^+$.
The fit is over 20 bins of 200 fs bin width. Thedata is background subtracted and includes the (very small) Monte Carlo
correction.}
\label{timefit}
\end{figure}
The mass plots for the $K\pi$ and $KK$ candidates used in this analysis are shown in
Figure~\ref{kkmass_lifetime}. The reflection background coming from $K\pi$ decays in the $KK$ mass distribution
is clearly seen and
the amount of the reflection is obtained by a mass fit to the signal sample. 
 The reflection 
mass shape is obtained from a high statistics Monte Carlo sample.
We assume that time evolution of the reflection is described by the lifetime of $K\pi$ and fit the 
reduced proper time distribution of the $K\pi$ and $KK$ samples at the same time.
There are four fit parameters: $K\pi$ lifetime, the lifetime difference between
$K\pi$ and $KK$ and two background levels for the $K\pi$ and $KK$.
The signal contributions for the $K^-\pi^+$,
 $K^-K^+$ and the reflection from misidentified 
 $D^0\rightarrow K^-\pi^+$ in the reduced proper time histograms are described by
$f(t^\prime){\rm exp}(-t^\prime/\tau)$ in the fit likelihood.
$f(t^\prime)$ is a function for any deviation from a pure exponential signal due to
acceptance and absorption variation. The background number parameters are either floated or fixed to the 
number of events in mass sidebands using a Poisson term, which ties the background level to that observed in the
sidebands, in the fit likelihood.
The bin width of the reduced proper time is 200 fs.
The fit results to the observed proper time distribution for $K\pi$ and $KK$ are shown in Figure~\ref{timefit}.
By changing the selection cuts and trying different fitting methods the systematic errors are
estimated. We tested the particle identification hypothesis for kaon candidates and the minimum detachment 
required between primary and secondary vertices. The former affects the level of reflection backgrounds 
and the latter affects the amount of non-charm backgrounds.
Since the results could be affected by various charm reflections which produce curved mass distributions
and are therefore not properly subtracted from symmetrically placed sidebands,
we check this effect by reducing sideband width by half. We also tried two different options of background handling
as stated in the previous paragraph.
The differences in fitted $y_{\rm cp}$ are added quadratically.
We tried other variations of selection and fitting and found that results are nearly identical to standard fits.
We obtained 


\begin{table*}[t]
\caption{CP asymmetry in D decays.}
\label{cpresult}
\begin{tabular}{|c|c|c|c| }
\hline
\raisebox{0pt}[16pt][6pt]{}&
$D^+\rightarrow K^-K^+\pi^+$      &
$D^0\rightarrow K^-K^+$ &
$D^0\rightarrow \pi^-\pi^+$ \\
\hline
\raisebox{0pt}[16pt][6pt]{FOCUS} &
$+0.006\pm 0.011\pm 0.005$      &
$-0.001\pm 0.022\pm 0.015$ &
$+0.048\pm 0.039\pm 0.025$ \\
\hline
\raisebox{0pt}[16pt][6pt]{E791} &
$-0.014\pm 0.029$  &
$-0.010\pm 0.049\pm 0.012$ &
$-0.049\pm 0.078\pm 0.025$ \\
\hline
\end{tabular}
\end{table*}

\begin{equation}
y_{\rm cp}=(3.42\pm 1.39\pm 0.74)\%
\label{eq:app2}
\end{equation}

Our result~\cite{FOCUSlifetime} is consistent with that of E791 measurement~\cite{E791lifetime} 
but the sign of our measurement is opposite to that of CLEO.~\cite{CLEOlifetime}
The theoretical expectation for the strong phase involved in this process 
varies~\cite{CLEOFOCUS} and caution is needed in combining the $y_{\rm cp}$ and the $y^\prime$(which is a rotational transformation of 
mixing parameters $x$ and $y$ that depends on a strong phase shift) into one mixing parameter, $y$.

\section{CP \rm{violation}}

It is well known that CP violating effects occur in a decay process only
if the decay amplitude is the sum of two different parts, whose phases are made of
 a weak and a strong contribution. 
The expected asymmetries are around $10^{-3}$.
We look at the Cabibbo suppressed decay modes which have the largest branching fractions and
select decay modes, $D^+\rightarrow K^-K^+\pi^+$, $D^0\rightarrow K^-K^+$ and
$D^0\rightarrow \pi^-\pi^+$. We use the sign of the bachelor pion in the 
$D^{*\pm}$ decay to tag the neutral D as either a $D^0$ or a $\bar D^0$.
In photoproduction fixed target experiment we must account for the different 
production rates of
charm particles and antiparticles. This is done using Cabibbo favored modes 
$D^0\rightarrow K^-\pi^+$ and $D^+\rightarrow K^-\pi^+\pi^+$. 
This way also has the advantage that most of the corrections
due to inefficiencies cancel out by dividing the rate of singly Cabibbo 
 suppressed decays by Cabibbo favored decay modes. 
 We assume that there is no measurable CP violation
in the Cabibbo favored decays. The CP asymmetry can be written as

\begin{equation}
A_{\rm cp}=\frac{\eta(D)-\eta(\bar D)}{\eta(D)+\eta(\bar D)}\!
\label{eq:app3}
\end{equation}
where $\eta$ is (considering for example the decay mode 
$D^0\rightarrow K^-K^+$)
\begin{equation}
\eta(D)=\frac{N(D^0\rightarrow K^-K^+)}{N(D^0\rightarrow K^-\pi^+)}\!
\label{eq:app4}
\end{equation} 
and $N(D^0\rightarrow K^-K^+)$ is the efficiency corrected number of candidate decays.
To tag the flavor of the neutral D meson $D^*$ tagging is required. 
As a result
the statistical error in the neutral decay modes are larger than those of 
charged decay mode. Our measurements~\cite{FOCUSCP} are summarized in the Table~\ref{cpresult} and results from
E791 experiment~\cite{E791CP} are also presented. 
The FOCUS measurement is 2$-$3 times better than the
the previous measurements by E791. We found no evidence for CP violation in 
the singly Cabibbo suppressed decay modes.

\end{document}